\input ./style/arxiv-general.cfg
\documentclass[aoas,MSNbibl,nameyear,seceqn,rotating,dvips]{arximspdf}
\makeatletter
   \@ifpackageloaded{graphicx}{}{\usepackage{graphicx}}
\makeatother
\usepackage{dcolumn}

\doi{10.1214/15-AOAS808}
\volume{9}
\issue{2}
\pubyear{2015}
\firstpage{640}
\lastpage{664}
\docsubty{FLA}

\makeatletter
\newcolumntype{d}[1]{D{.}{.}{#1}}
\def\sfrac#1#2{#1/#2}
\def\vfrac#1#2{(#1)/#2}

\def\sklvfrac#1#2{((#1)/#2)}

\newcommand{\rrvert}{\vert}
\newcommand{\rrVert}{\Vert}
\newcommand{\llvert}{\vert}
\newcommand{\llVert}{\Vert}
\renewcommand{\mid}{|}
\makeatother

\begin{document}
\begin{frontmatter}

\title{Bayesian group Lasso for nonparametric varying-coefficient
models with application to functional genome-wide association studies}
\runtitle{Bayesian group Lasso and functional GWAS}

\begin{aug}
\author[A]{\fnms{Jiahan}~\snm{Li}\thanksref{M1}\ead[label=e1]{jli7@nd.edu}},
\author[B]{\fnms{Zhong}~\snm{Wang}\thanksref{M2,T2}\ead[label=e2]{zhongwang@bjfu.edu.cn}},
\author[C]{\fnms{Runze}~\snm{Li}\thanksref{M3,T3}\ead[label=e3]{rzli@psu.edu}}
\and
\author[D]{\fnms{Rongling}~\snm{Wu}\corref{}\thanksref{M3,M2,T4}\ead[label=e4]{rwu@hes.hmc.psu.edu}\ead[label=e5]{rwu@bjfu.edu.cn}}
\runauthor{Li, Wang, Li and Wu}
\affiliation{University of Notre Dame\thanksmark{M1}, Beijing Forestry
University\thanksmark{M2} and  Pennsylvania State
University\thanksmark{M3}}
\address[A]{J. Li\\
Department of Applied and Computational\\
\quad Mathematics and Statistics \\
University of Notre Dame\\
Notre Dame, Indiana 46556\\
USA\\
\printead{e1}}
\address[B]{Z. Wang\\
Center for Computational Biology\\
Beijing Forestry University \\
Beijing\\
China 100083\\
\printead{e2}}
\address[C]{R. Li\\
Department of Statistics \\
The Methodology Center\\
Pennsylvania State University \\
University Park, Pennsylvania 16802\\
USA\\
\printead{e3}}
\address[D]{R. Wu\\
Center for Statistical Genetics\\
Pennsylvania State University\\
Hershey, Pennsylvania 17033\\
USA\\
and\\
Center for Computational Biology\\
Beijing Forestry University\\
Beijing\\
China 10008\\
\printead{e4}}
\end{aug}
\thankstext{T2}{Supported in part by NSFC Grant 31470675.}
\thankstext{T3}{Supported in part by NIDA Grant P50-DA10075.}
\thankstext{T4}{Supported in part by Grants U01 HL119178,
NIH/NHLBI-1U10HL098115 and UL1 TR000127.}

%
\received{\smonth{12} \syear{2012}}
%
\revised{\smonth{1} \syear{2015}}

%
\begin{abstract}
Although genome-wide association studies (GWAS) have proven powerful
for comprehending the genetic
architecture of complex traits, they are challenged by a high dimension
of single-nucleotide polymorphisms (SNPs) as predictors, the presence
of complex environmental factors, and longitudinal or functional
natures of many complex traits or diseases. To address these
challenges, we propose a high-dimensional varying-coefficient model for
incorporating functional aspects of phenotypic traits into GWAS to
formulate a so-called functional GWAS or \textit{f}GWAS. The Bayesian
group lasso and the associated MCMC algorithms are developed to
identify significant SNPs and estimate how they affect longitudinal
traits through time-varying genetic actions. The model is generalized
to analyze the genetic control of complex traits using subject-specific
sparse longitudinal data. The statistical properties of the new model
are investigated through simulation studies. We use the new model to
analyze a real GWAS data set from the Framingham Heart Study, leading
to the identification of several significant SNPs associated with
age-specific changes of body mass index. The \textit{f}GWAS model,
equipped with the Bayesian group lasso, will provide a useful tool for
genetic and developmental analysis of complex traits or diseases.
\end{abstract}

%
\end{frontmatter}

\setcounter{footnote}{3}

\section{Introduction}\label{sec1}
Phenotypic traits of paramount importance to agriculture and human
health are quantitatively inherited, involving an unknown (usually very
high) number of genes and undergoing a series of developmental pathways
and events [\citet{LynWal98}; \citet{WuLin06}]. These
complexities have made the genetic analysis of quantitative traits one
of the most difficult tasks in biological sciences. Recently emerging
genome-wide association studies (GWAS) have provided a great promise to
systematically characterize the genetic control of complex traits and
have been increasingly instrumental for the identification of
significant genetic variants that control phenotypic variation
[\citet{Shual09}; \citet{Taketal09}; \citet{Teietal09}; \citet{Yanetal10}]. In human genetics, these results have
started to gain a growing body of novel findings with potential
clinical relevance [\citet{Dal10}]. In plant and animal genetics, GWAS,
with the advent of a continuously falling genotyping cost, have been
considered more seriously than any time before [\citet{FilMal12}]. Despite their powerful impact on genetic studies, however,
GWAS also encounter tremendous challenges from statistical analysis and
interpretation.

First, GWAS usually genotype hundreds of thousands of single-nucleotide
polymorphisms (SNPs) on thousands of subjects, leading to a number of
SNPs strikingly larger than the sample size used. Thus, to analyze
these SNPs, simple univariate linear regression has to be used for
individual tests. However, this method ignores the effects of other
SNPs while assessing one particular SNP, and is subjected to a severe
adjustment issue for multiple comparisons. Moreover, in biology and
biomedicine, a phenotypic trait can always be better described by a
dynamic trajectory because the trait undergoes a developmental process
[\citet{WuLin06}]. For example, human body height growth is a process
from infancy to adulthood; the genetic study of adult height only, as
conducted in many current GWAS [\citet{Let11}], provides limited
information about the developmental genetics of height and its
relationship with physical and mental characteristics at various stages
of growth. In clinical trials, longitudinal measures are one of the
most common data types, including HIV dynamics, cancer growth and drug
response to varying doses [\citet{Wanetal09}]. In this article, we
address these issues by developing novel statistical models and
algorithms that can analyze multiple SNPs simultaneously and integrate
the developmental mechanisms of trait formation into a general GWAS
framework through mathematical functions. The extension of the models
to tackle genotype-environment interactions using GWAS is straightforward.

In a linear regression model for GWAS where SNPs are predictors,
multiple regression breaks down when the number of predictors far
exceeds the number of subjects. Alternatively, variable selection
approaches could identify important genetic factors and enhance the
predictive power of the final model. For example, in analyzing
case-control cohorts, lasso regression [\citet{Tib96}] and
elastic-net regression [\citet{ZouHas05}] were studied by \citet{Wuetal09} and \citet{Choetal09}, respectively. \citet{Lietal12} and \citet{HeLin11} further proposed two-stage variable selection approaches
to identify disease susceptibility genes. These methods, however, are
restricted to models with a single phenotypic measurement from each subject.

For genetic studies of dynamic traits that are measured repeatedly at
multiple time points, \citet{WuLin06} proposed a conceptual model
called functional mapping by incorporating longitudinal and functional
data analysis into a genetic design. Depending on the availability of
explicit mathematical equations to describe a biological process,
functional mapping uses parametric, nonparametric or semiparametric
approaches for modeling nonlinear effects of genetic variants over time
and further revealing a dynamic landscape of interplay between genes
and developmental pattern. \citet{Dasetal11} implemented functional
mapping into a GWAS setting, leading to the birth of a so-called
functional GWAS or \textit{f}GWAS model. The basic principle of functional
mapping and \textit{f}GWAS is to model and predict the temporal pattern of
genetic effects on a particular trait or disease in a quantitative
manner. Time-varying change of gene expression has been found to be a
ubiquitous phenomenon because different metabolic pathways, regulated
by genes directly or indirectly, are required for an organism to best
adapt to developmental alteration. In a genetic study of body mass
index (BMI) by linkage mapping, \citet{Goretal03} identified
different BMI susceptibility genes as well as different modes of
inheritance triggered by these genes in children and adults. A common
variant in the obesity-associated FTO gene, identified by a genome-wide
search, was observed to be reproducibly associated with BMI and obesity
from childhood into old age, but displayed varying magnitudes of
genetic effects between child and adult stages [Fraying et~al. (\citeyear{F2007})].

To increase its applicability in clinical genomics, \textit{f}GWAS could
further accommodate irregular longitudinal data measured at
subject-specific time points. But both functional mapping and \textit{f}GWAS analyze SNPs individually or pairwise, and are incapable of
depicting a comprehensive picture of the genetic architecture of
dynamic traits. The motivation of this article is to develop a variable
selection model for \textit{f}GWAS, with a focus on nonparametric modeling
of temporal genetic effects of SNPs. Variable selection in a
nonparametrical setting is equivalent to selecting a subset of
predictors with nonzero functional coefficients. \citet{LinZha06}
developed COSSO for model selection in a smoothing spline ANOVA model,
with the penalty term being the sum of component norms. \citet{ZhaLin06} further extended it to nonparametric regression in an
exponential family. \citet{WanLiHua08} estimated time-varying effects
using basis expansion and selected significant predictors by imposing
SCAD penalty functions on the $L_2$-norm of these basis expansions.

We propose a Bayesian group lasso approach for variable selections in
nonparametric varying-coefficient models.
Group lasso was first
proposed by \citet{YuaLin06}. They considered the problem of
selecting important groups of independent variables in linear
regression models and generalized lasso by encouraging sparsity at the
group level. However, since the Hessian is not defined at the optimal
solution, they did not provide variance estimates for the regression
coefficients. Here, we express time-varying effects as a linear
combination of Legendre polynomials, and in such a case, the selection
of important predictors corresponds to the selection of groups of
polynomials. We develop a Bayesian hierarchical model for group
variable selection and estimate all parameters by MCMC algorithms. Our
method provides not only point estimates but also interval estimates of
all parameters, and the traditional Bayesian lasso [\citet{ParCas08}] is its special case in which the response variable is univariate.

In Section~\ref{sec2}, we introduce the \textit{f}GWAS model that connects
genotypes and irregular longitudinal phenotypical data. Section~\ref{sec3} shows
the Bayesian hierarchical representation for this nonparametric
varying-coefficient model, where group lasso penalties are applied to
individual functional coefficients. The posterior computations as well
as the interpretation of the results are described in Section~\ref{sec4}. In
Section~\ref{sec5}, the statistical properties of the model are investigated
through simulation studies. Section~\ref{sec6} provides the application to a
real GWAS example from the Framingham Heart Study that analyzes
age-specific changes of genetic effects on body mass index (BMI). BMI
is a heuristic measure of body weight based on a person's weight and
height, providing the most widely used diagnostic tool to identify
whether individuals are underweighted, overweighted or obese, and,
further, to examine their risk of developing obesity-related diseases,
such as hypertension, type~2 diabetes and cardiovascular diseases
[\citet{Fra07}]. We use a nonparametric approach based on orthogonal
polynomials to approximate age-specific change in BMI. The discussion
about the new model is given in Section~\ref{sec7}.\looseness=1

\section{The \textit{f}GWAS model}\label{sec2}
The model for functional genome-wide association studies (\textit{f}GWAS)
is the integration of functional data analysis and genome-wide
association studies. The primary goal of the \textit{f}GWAS is to study
the dynamic pattern of genetic actions and interactions triggered by
significant SNPs throughout the entire genome. Beyond traditional GWAS,
\textit{f}GWAS targets phenotypic traits that are measured longitudinally
at repeated time points. Suppose in a genome-wide association study
involving $n$ subjects, a continuous longitudinal trait of interest is
measured at irregularly spaced time points, which are not common to all
subjects. Let ${\mathbf y}_i=(y_i(t_{i1}),\ldots,y_i(t_{iT_i}))^T$ be
the $T_i$-dimensional vector of measurements on subject $i$ where
${\mathbf t}_i=(t_{i1},\ldots,t_{iT_i})^T$ is the corresponding vector
of measurement time points after standardization. ${\mathbf
y}_i$ can be described as
\begin{eqnarray}
\nonumber
\pmatrix{y_i(t_{i1})
\vspace*{3pt}\cr
\vdots
\vspace*{3pt}\cr
y_i(t_{iT_i})}&=&\pmatrix{\mu(t_{i1})
\vspace*{3pt}\cr
\vdots
\vspace*{3pt}\cr
\mu(t_{iT_i})}+ \pmatrix{\alpha_1(t_{i1})&\cdots&
\alpha_{q}(t_{i1})
\vspace*{3pt}\cr
\vdots&&\vdots
\vspace*{3pt}\cr
\alpha_1(t_{iT_i})&
\cdots&\alpha_{q}(t_{iT_i})} \pmatrix{X_{i1}
\vspace*{3pt}\cr
\vdots
\vspace*{3pt}\cr
X_{iq}}\nonumber
\\
&&{}+\pmatrix{a_1(t_{i1})&\cdots&a_{p}(t_{i1})
\vspace*{3pt}\cr
\vdots&&\vdots
\vspace*{3pt}\cr
a_1(t_{iT_i})&\cdots&a_{p}(t_{iT_i})}
\pmatrix{\xi_{i1}
\vspace*{3pt}\cr
\vdots
\vspace*{3pt}\cr
\xi_{ip}}
\\
&&{}+\pmatrix{d_1(t_{i1})&\cdots&d_{p}(t_{i1})
\vspace*{3pt}\cr
\vdots&&\vdots
\vspace*{3pt}\cr
d_1(t_{iT_i})&\cdots&d_{p}(t_{iT_i})}
\pmatrix{\zeta_{i1}
\vspace*{3pt}\cr
\vdots
\vspace*{3pt}\cr
\zeta_{ip}}+
\pmatrix{e_i(t_{i1})
\vspace*{3pt}\cr
\vdots
\vspace*{3pt}\cr
e_i(t_{iT_i})}.\nonumber
\end{eqnarray}
We introduce matrix notation for a succinct presentation. Let
$\bolds{\alpha}(t_{i\ell})=(\alpha_1(t_{i\ell}),\ldots,\alpha_q(t_{i\ell}))^T$
be the $q$-dimensional vector of covariate effects, ${\mathbf
X}_i=(X_{i1},\ldots,X_{iq})^T$ be the observed covariate vector for subject
$i$, ${\mathbf a}(t_{i\ell})=\break (a_1(t_{i\ell}),\ldots,a_p(t_{i\ell}))^T$
and  ${\mathbf
d}(t_{i\ell})=(d_1(t_{i\ell}),\ldots,d_p(t_{i\ell}))^T$ be the
$p$-dimensional vectors of the additive and dominant effects of SNPs,
respectively. Furthermore, let $\bolds{\xi}_{i}=(\xi_{i1},\ldots,\xi
_{ip})^T$ and
$\bolds{\zeta}_i=(\zeta_{i1},\ldots,\zeta_{ip})^T$ be the indicator
vectors of the
additive and dominant effects of SNPs for subject $i$. Thus, at time point
$t_{i\ell}$,
%
\begin{eqnarray}
\label{eqregressionGWAS1}
\nonumber
y_i(t_{i\ell})&=&
\mu(t_{i\ell})+\bolds{\alpha}(t_{i\ell})^T{\mathbf
X}_i+{\mathbf a}(t_{i\ell})^T\bolds{
\xi}_{i}+{\mathbf d}(t_{i\ell})^T\bolds{
\zeta}_{i}+e_i(t_{i\ell}),
\nonumber\\[-8pt]\\[-8pt]
\eqntext{i=1,\ldots,n, \ell=1,\ldots,T_i,}
\end{eqnarray}
where $\mu(t_{i\ell})$ is the overall mean and $e_i(t_{i\ell})$ is the
residual error assumed to follow a $\mathrm{N}(0,\sigma^2(t_{i\ell}))$
distribution. The $j$th elements of $\bolds{\xi}_{i}$ and $\bolds{\zeta
}_{i}$ are
defined as
\begin{eqnarray*}
\xi_{i,j} &=& \cases{ 1, &\quad if the genotype of SNP $j$ is $AA$,
\vspace*{3pt}\cr
0, &
\quad if the genotype of SNP $j$ is $Aa$,
\vspace*{3pt}\cr
-1,&\quad if the genotype of SNP $j$ is $aa$,}
\\
\zeta_{i,j} &=& \cases{ 1, &\quad if the genotype of SNP $j$ is $Aa$,
\vspace*{3pt}\cr
0,
&\quad if the genotype of SNP $j$ is $AA$ or $aa$.}
\end{eqnarray*}
In other words, $a_j(t_{i\ell})$ represents the average effect of
substituting one allele for the other, and $d_j(t_{i\ell})$ represents
how the average genotypic value of the heterozygote deviates from the
mean of the homozygotes.

In the \textit{f}GWAS model, the effects of covariates and SNPs are
assumed to be functions of time. Many methods of estimating
time-varying coefficients of a linear model in a longitudinal data
setting have been proposed and studied, including basis expansion
methods, local polynomial kernel methods and smoothing spline methods.
Among these techniques, Legendre polynomials have been widely used by
quantitative geneticists for modeling the growth curves [\citet{LinWu06}], the programmed cell death (PCD) process [\citet{Cuietal08}] or
the genetic effects responsible for other traits [e.g., \citet{SucSzy11}; \citet{YanXu07}; \citet{Dasetal11}]. By approximating
time-varying effects using Legendre polynomials, the expansion
coefficients can be solved through regression. Moreover, the biological
evidence or the prior belief about the time-dependency of genetic
control can be integrated by just truncating the series. Motivated by
these studies, we approximate the effect of the $k$th covariate by a
Legendre polynomial of order $v-1$:
%
\begin{equation}
\bigl(\alpha_k(t_{i1}),\ldots,\alpha_k(t_{iT_i})
\bigr)^T=U_i{\mathbf r}_k,\qquad k=1,\ldots,q,
\end{equation}
where ${\mathbf r}_k=(r_{k0},\ldots,r_{k(v-1)})^T$ are the Legendre
polynomial coefficients, and
%
\begin{eqnarray}
U_i=\pmatrix{{\mathbf u}^T_{i1}
\vspace*{3pt}\cr
\vdots
\vspace*{3pt}\cr
{\mathbf
u}^T_{iT_i}}=\pmatrix{1&t_{i1} &\displaystyle\tfrac{1}{2}
\bigl(3t_{i1}^2-1\bigr)&\cdots
\vspace*{3pt}\cr
\vdots&\vdots&\vdots&
\vdots
\vspace*{3pt}\cr
1&t_{iT_i}&\displaystyle\tfrac{1}{2}\bigl(3t_{iT_i}^2-1
\bigr)& \cdots}
\end{eqnarray}
are Legendre polynomial functions. Similarly, other time-varying
effects can be represented as
%
\begin{eqnarray}
\bigl(a_j(t_{i1}),\ldots,a_j(t_{iT_i})
\bigr)^T&=&U_i{\mathbf b}_j,\qquad j=1,\ldots,p,
\\
\bigl(d_j(t_{i1}),\ldots,d_j(t_{iT_i})
\bigr)^T &=& U_i{\mathbf c}_j,\qquad j=1,\ldots,p,
\\
\bigl(\mu(t_{i1}),\ldots,\mu(t_{iT_i}) \bigr)^T &=& U_i{
\mathbf m},
\end{eqnarray}
where ${\mathbf b}_j=(b_{j0},\ldots,b_{j(v-1)})^T$ are the Legendre
polynomial coefficients for the additive effect of the $j$th SNP,
${\mathbf c}_j=(c_{j0}, \ldots,c_{j(v-1)})^T$ are the Legendre
polynomial coefficients for the dominant effect of the $j$th SNP, and
${\mathbf m}=(m_{0},\ldots,m_{v-1})^T$ are the Legendre polynomial
coefficients for the overall mean function.

After introducing Legendre polynomials to approximate time-varying
effects of covariates and SNPs, the full model of \textit{f}GWAS becomes
%
\begin{eqnarray}
\label{eqregressionGWAS} y_i(t_{i\ell})&=&{\mathbf u}^T_{il}{
\mathbf m}+\bigl({\mathbf u}^T_{il}{\mathbf r}_1,\ldots,{
\mathbf u}^T_{il}{\mathbf r}_q\bigr){\mathbf
X}_i\nonumber
\\
&&{} +\bigl({\mathbf u}^T_{il}{\mathbf
b}_1,\ldots,{\mathbf u}^T_{il}{\mathbf
b}_p\bigr)\bolds{\xi}_{i}
+\bigl({\mathbf u}^T_{il}{\mathbf c}_1,\ldots,{\mathbf
u}^T_{il}{\mathbf c}_p\bigr)\bolds{
\zeta}_{i}+e_i(t_{i\ell}),
\\
\eqntext{i=1,\ldots,n, \ell=1,\ldots,T_i.}
\end{eqnarray}

Last, since measurements within each subject are possibly correlated
with one another, we assume that ${\mathbf e}_i=(e_i(t_{i1}),\ldots,
e_i(t_{iT_i}))^T$ follows a multivariate normal distribution with zero
mean and covariance matrix $\Sigma_i$. Both parametric and
nonparametric methods have been developed to model the structure of
covariance between longitudinal measurements [\citet{MaCasWu02}; \citet{Zhaetal05}; \citet{YapFanWu09}]. In particular, we employ the
first-order autoregressive [AR(1)] model to approximate the residual
covariance matrix. This covariance structure allows different
measurement time points for different subjects, and assumes a constant
variance over time and an exponentially decaying correlation, $\rho
^{|t_{i2}-t_{i1}|}, 0 < \rho< 1$, between two measurements. Moreover,
the matrix determinant in the likelihood function can be easily
computed. In our real data example, the variance of repeated
measurements is stable over time. In longitudinal data sets with
variance heteroscedasticity, however, a Transform-Both-Sides (TBS)
technique [\citet{Wuetal04}] can be employed to satisfy the variance
stationarity assumption in the AR(1) model.

\section{Bayesian hierarchical representation for group Lasso penalties}\label{sec3}
In high-dimensional regression problems, such as GWAS, a regularized
approach is preferred to identify predictors with nonzero effects and
to achieve better out-of-sample predictive performance. When parameters
that we would like to penalize are finite-dimensional, we may apply
different penalty functions to them to perform variable selection. But
when these parameters are nonparametric smooth functions, a traditional
regularization procedure cannot be directly applied. In this situation,
regularized estimation for selecting important predictors is equivalent
to selecting functional coefficients that are not identically zero.

Let $\llVert {\mathbf b}_j\rrVert $ be the $L_2$ norm of the vector
${\mathbf b}_j$. The time-varying additive effect of the $j$th SNP is
identically zero if and only if $\llVert {\mathbf b}_j\rrVert =0$.
Therefore, if we estimate additive effects by a Legendre polynomial of
order $v$, and would like to identify significant additive effects via
penalized methods, we could partition all parameters of additive\vspace*{1pt}
effects $({\mathbf b}^T_1, \ldots, {\mathbf b}^T_p)$ into $p$ groups of
size $v$ according to $p$ SNPs, and encourage sparse solution at the
group level or select a subset of groups with nonzero $L_2$ norms. That
is, the group lasso minimizes the following penalized least square:
%
\begin{equation}
\label{eqPenalizedLSlasso}\frac{1}{2}\llVert{\mathbf y}-{\bolds\mu
}\rrVert^2+
\lambda\sum_{j=1}^p\llVert{\mathbf
b}_j\rrVert+\lambda^*\sum_{j=1}^p
\llVert{\mathbf c}_j\rrVert,
\end{equation}
where ${\mathbf y}^T=({\mathbf y}^T_1,\ldots,{\mathbf y}^T_n)$,
${\bolds\mu}^T=E {\mathbf y}^T=({\bolds\mu}^T_1,\ldots,{\bolds\mu
}^T_n)$ and $\lambda$ and $\lambda^*$ are two regularization
parameters. $\lambda$ and $\lambda^*$ control the amount of shrinkage
toward zero: the larger their values, the greater the amount of
shrinkage. They should be adaptively determined from the data to
minimize an estimate of expected prediction error.

From a Bayesian perspective, the group lasso estimates can be
interpreted as posterior mode estimates when the regression parameters
have multivariate independent and identical Laplace priors. Therefore,
when group lasso penalties are imposed on the Legendre coefficients of
additive and dominant effects, the conditional prior for ${\mathbf
b}_j$ is a multivariate Laplace distribution with the scale parameter
$(v\lambda^2/\sigma^2)^{-1/2}$:
%
\begin{equation}
\label{eqlaplaceb}\pi\bigl({\mathbf b}_j\mid\sigma^2\bigr)=
\bigl(v\lambda^2/\sigma^2\bigr)^{v/2}
\operatorname{exp}\bigl(-\bigl(v\lambda^2/\sigma^2
\bigr)^{-1/2}\llVert{\mathbf b}_j\rrVert\bigr),
\end{equation}
and the conditional multivariate Laplace prior for dominant effect
${\mathbf c}_j$ is
%
\begin{equation}
\label{eqlaplacec}\pi\bigl({\mathbf c}_j\mid\sigma^2\bigr)=
\bigl(v\lambda^{*2}/\sigma^2\bigr)^{v/2}
\operatorname{exp}\bigl(-\bigl(v\lambda^{*2}/\sigma^2
\bigr)^{-1/2} \llVert{\mathbf c}_j\rrVert\bigr).
\end{equation}

To ensure the derived conditional distribution of ${\mathbf b}_j$ has a
standard form, we rewrite the multivariate Laplace prior distribution
as a scale mixture of a multivariate Normal distribution with a Gamma
distribution, that is,
\begin{eqnarray*}
\nonumber
&&\mbox{M-Laplace}\bigl({\mathbf b}_j\mid0,\bigl(v
\lambda^2/\sigma^2\bigr)^{-1/2}\bigr)
\\
\nonumber
&&\qquad \propto \bigl(v\lambda^2/\sigma^2
\bigr)^{v/2} \operatorname{exp}\bigl(-\bigl(v\lambda^2/
\sigma^2\bigr)^{1/2}\llVert{\mathbf b}_j\rrVert
\bigr)
\\
&&\qquad \propto\int_0^{\infty}\operatorname{MVN} \bigl({\mathbf
b}_j\mid{\mathbf0},\operatorname{diag}\bigl(\sigma^2
\tau_j^2,\ldots,\sigma^2\tau_j^2
\bigr) \bigr) \operatorname{Gamma} \biggl(\tau_j^2 \bigg|
\frac{v+1}{2},\frac{2}{v\lambda^2} \biggr)\,d\tau_j^2,
\end{eqnarray*}
where $(v\lambda^2/\sigma^2)^{-1/2})$ is the scale parameter of
the multivariate Laplace distribution, a $v$-by-$v$ diagonal matrix
$\operatorname{diag}(\sigma^2\tau_j^2,\ldots,\sigma^2\tau_j^2)$ is the
covariance matrix of the multivariate normal distribution with mean zero,
$\frac{v+1}{2}$ is the shape parameter of the Gamma distribution, and
$\frac{2}{v\lambda^2}$ is the scale parameter of the Gamma
distribution. After integrating out $\tau^2_j$, the\vspace*{1pt} conditional prior
on ${\mathbf b}_j$ has the desired form (\ref{eqlaplaceb}). Then, in a
Bayesian hierarchical model, we can rewrite the multivariate Laplace
priors on ${\mathbf b}_j$ as
\begin{eqnarray*}
\nonumber
{\mathbf b}_j\mid\tau_j^2,
\sigma^2&\sim& \operatorname{MVN} \bigl({\mathbf0},\operatorname{diag}\bigl(
\sigma^2\tau_j^2,\ldots,\sigma^2
\tau_j^2\bigr) \bigr),
\\
\tau_j^2\mid\lambda&\sim&\operatorname{Gamma} \biggl(
\frac{v+1}{2},\frac{2}{v\lambda^2} \biggr).
\end{eqnarray*}
Likewise, the multivariate-Laplacian prior on ${\mathbf c}_j$ can be
replaced by
\begin{eqnarray*}
\nonumber
{\mathbf c}_j\mid\tau_j^{*2},
\sigma^2&\sim& \operatorname{MVN} \bigl({\mathbf0},\operatorname{diag}\bigl(
\sigma^2\tau_j^{*2},\ldots,\sigma^2
\tau_j^{*2}\bigr) \bigr),
\\
\tau_j^{*2} \mid\lambda&\sim&\operatorname{Gamma}
\biggl( \frac{v+1}{2},\frac{2}{v\lambda^{*2}} \biggr).
\end{eqnarray*}
Then, given $\lambda$ and $\lambda^*$, we have the following
hierarchical representation of the penalized regression model:
\begin{eqnarray*}
\nonumber
{\mathbf y}\mid{\mathbf m},{\mathbf r}_k,{\mathbf b}_j,{\mathbf
c}_j,\rho,\sigma^2 &\propto&(2\pi)^{-\vfrac{\sum_i^n T_i}{2}}\Biggl(
\prod_i^n \llvert
\Sigma_i\rrvert^{-1/2}\Biggr)e^{-1/2\sum_i^n
({\mathbf y}_i-{\bolds\mu}_i)^T\Sigma_i^{-1}({\mathbf y}_i-{\bolds\mu}_i)},
\\
\nonumber
{\mathbf m}&\sim&N_{v}(0,\Sigma_{m0}),
\\
\nonumber
{\mathbf r}_k&\sim&N_{v}(0,\Sigma_{r0}),
\qquad k=1,\ldots,q,
\\
\nonumber
{\mathbf b}_j|\tau_j^2,
\sigma^2&\sim&\operatorname{MVN} \bigl({\mathbf0},\operatorname{diag}\bigl(
\sigma^2\tau_j^2,\ldots,\sigma^2
\tau_j^2\bigr) \bigr),\qquad j=1,\ldots,p,
\\
\nonumber
\tau_j^2|\lambda&\sim&\operatorname{Gamma}
\biggl(\frac{v+1}{2},\frac{2}{v\lambda^2} \biggr),\qquad j=1,\ldots,p,
\\
\nonumber
{\mathbf c}_j|\tau_j^{*2},
\sigma^2&\sim& \operatorname{MVN} \bigl({\mathbf0},\operatorname{diag}\bigl(
\sigma^2\tau_j^{*2},\ldots,\sigma^2
\tau_j^{*2}\bigr) \bigr),\qquad j=1,\ldots,p,
\\
\nonumber
\tau_j^{*2}|\lambda^*&\sim&\operatorname{Gamma}
\biggl(\frac{v+1}{2},\frac{2}{v\lambda^{*2}} \biggr),\qquad j=1,\ldots,p,
\\
\nonumber
\rho&\sim&\mathrm{U}(-1,1),
\\
\nonumber
\sigma^2&\sim&\pi\bigl(\sigma^2\bigr),
\\
\sigma^2,\lambda,\lambda^*&>&0,
\end{eqnarray*}
where $\lambda$ and $\lambda^*$ are regularization parameters or group
lasso parameters that control the shrinkage intensities in estimating
genetic effects. We assign a conjugate multivariate normal prior to
${\mathbf m}$ when estimating the overall mean function. We also assign
conjugate multivariate normal priors to the Legendre coefficients of
covariates ${\mathbf r}_k,k=1,\ldots,q$, because covariates in GWAS are
usually low dimensional and are not the target of variable selection.
We assume a Uniform prior on $[-1,1]$ for~$\rho$, the autoregressive
parameter in the assumed AR(1) covariance matrix. Finally, since the
data are usually sufficient to estimate $\sigma$, we can use a
noninformative prior such as $\pi(\sigma^2)=1/\sigma^2$ for $\sigma^2$.\looseness=-1

Traditionally, two group lasso parameters $\lambda$ and $\lambda^*$ can
be prespecified by cross-validation or generalized cross-validation.
However, in the Bayesian group lasso setting, $\lambda$ and $\lambda^*$
can be estimated along with other parameters by assigning appropriate
hyperpriors to them. This procedure determines the amount of
regularization from the data and avoids refitting the model. In
particular, the following conjugate gamma priors are considered,
\[
\pi\biggl(\frac{\lambda^2}{2} \biggr)\sim\operatorname{Gamma}(a,b)
\quad\mbox{and}\quad\pi\biggl(\frac{\lambda^{*2}}{2} \biggr)\sim
\operatorname{Gamma}
\bigl(a^*,b^*\bigr),
\]
where $a$, $b$, $a^*$ and $b^*$ are small values so that the priors are
essentially noninformative. With this specification, group lasso
parameters can simply join the other parameters in the Gibbs sampler.

\section{Posterior computation and interpretation}\label{sec4}
We estimate the unknown parameters and hyperparameters by sampling from
their conditional posterior distributions through MCMC algorithms.
Given the data likelihood and prior distributions, the posterior
distributions of all unknowns can be obtained by Bayes' \mbox{theorem}. For
most of the parameters, the conditional posterior distributions have
closed forms by conjugacy, which facilitates drawing posterior samples.

Assuming that priors for different predictors are independent, we can
express the joint posterior distribution of all parameters as
\begin{eqnarray*}
\nonumber
&&\pi\bigl({\mathbf m},{\mathbf r}_k,{\mathbf b}_j,
\tau^2_j,\lambda,{\mathbf c}_j,
\tau^{*2}_j,\lambda^*,\sigma^2,\rho|{\mathbf y}
\bigr)
\\
\nonumber
&&\qquad \propto\pi({\mathbf y}|\cdot)\pi({\mathbf m})\pi\bigl(\sigma^2
\bigr)\pi(\rho)\prod_{k=1}^q\pi({\mathbf
r}_k)
\\
&&\quad\qquad{}\times \prod_{j=1}^p \pi\bigl({\mathbf
b}_j|\tau_j^2\bigr) \pi\bigl(
\tau_j^2|\lambda\bigr)\pi(\lambda)\pi\bigl({\mathbf
c}_j|\tau_j^{*2}\bigr)\pi\bigl(
\tau_j^{*2}|\lambda^*\bigr)\pi\bigl(\lambda^*\bigr).
\end{eqnarray*}

Conditional on the parameters $({\mathbf r}_k,{\mathbf b}_j,\tau
^2_j,\lambda,{\mathbf c}_j,\tau^{*2}_j,\lambda^*,\sigma^2,\rho)$, we
derive the conditional posterior distribution of ${\mathbf m}$ as
\begin{eqnarray*}
\nonumber
&&\pi\bigl({\mathbf m}|{\mathbf y},{\mathbf r}_k,{\mathbf
b}_j,\tau^2_j,\lambda,{\mathbf
c}_j,\tau^{*2}_j,\lambda^*,
\sigma^2,\rho\bigr)
\\
\nonumber
&&\qquad \propto\pi({\mathbf m})\pi({\mathbf y|\cdot})
\\
\nonumber
&&\qquad \propto\operatorname{exp} \Biggl(-\frac{1}{2}{\mathbf
m}^T \Sigma_{m0}^{-1}{\mathbf m}
\\
&&\hspace*{53pt}{} -\frac{1}{2}
\sum_{i=1}^n({\mathbf y}_i - {\bolds
\mu}_{i(-m)}-U_i{\mathbf m})^T\Sigma_i^{-1}({
\mathbf y}_i-{\bolds\mu}_{i(-m)}-U_i{\mathbf m}) \Biggr)
\\
\nonumber
&&\qquad \propto\operatorname{exp} \Biggl({\mathbf m}^T
\Sigma_{m0}^{-1}{\mathbf m} +\sum_{i=1}^n(U_i{
\mathbf m})^T\Sigma_i^{-1}(U_i{\mathbf
m})
\\
&&\hspace*{61pt}{}-2\sum_{i=1}^n ({\mathbf
y}_i-{\bolds\mu}_{i(-m)})^T\Sigma_i^{-1}(U_i{
\mathbf m}) \Biggr)
\\
&&\qquad \propto\operatorname{exp} \Biggl({\mathbf m}^T\Biggl(
\Sigma_{m0}^{-1}+\sum_{i=1}^n
U_i^T\Sigma_i^{-1}U_i
\Biggr){\mathbf m}-2\sum_{i=1}^n ({\mathbf
y}_i - {\bolds\mu}_{i(-m)})^T\Sigma_i^{-1}(U_i{
\mathbf m}) \Biggr).
\end{eqnarray*}
Hence, the conditional posterior distribution of ${\mathbf m}$ is
$\operatorname{MVN}_v({\bolds\mu}_{m},\Sigma_{m})$, where
\[
{\bolds\mu}_{m}= \Biggl(\Sigma_{m0}^{-1}+\sum
_{i=1}^n U_i^T
\Sigma_i^{-1}U_i \Biggr)^{-1}
\Biggl(\sum_{i=1}^n ({\mathbf
y}_i-{\bolds\mu}_{i(-m)})^T\Sigma_i^{-1}U_i
\Biggr)^T,
\]
and
\[
\Sigma_{m}= \Biggl(\Sigma_{m0}^{-1}+\sum
_{i=1}^n U_i^T
\Sigma_i^{-1}U_i \Biggr)^{-1}.
\]

Similarly, since ${\mathbf r}_k$, ${\mathbf b}_j$ and ${\mathbf c}_j$
have conjugate multivariate normal priors, the posterior distribution
for ${\mathbf r}_k$ is $\operatorname{MVN}_v({\bolds\mu}_{r_k},\Sigma
_{r_k})$, with
\begin{eqnarray*}
{\bolds\mu}_{r_k} &=& \Biggl(\Sigma_{r0}^{-1}+\sum
_{i=1}^n (X_{ik}U_i)^T
\Sigma_i^{-1}(X_{ik}U_i)
\Biggr)^{-1}
\\
&&{}\times  \Biggl( \sum_{i=1}^n
({\mathbf y}_i-{\bolds\mu}_{i(-r_k)})^T
\Sigma_i^{-1}(X_{ik}U_i)
\Biggr)^T,
\end{eqnarray*}
and
\[
\Sigma_{r_k}= \Biggl(\Sigma_{r0}^{-1}+\sum
_{i=1}^n (X_{ik}U_i)^T
\Sigma_i^{-1}(X_{ik}U_i)
\Biggr)^{-1},
\]
the posterior distribution for ${\mathbf b}_j$ is $\operatorname{MVN}_v({\bolds\mu}_{b_j},\Sigma_{b_j})$, with
\begin{eqnarray*}
{\bolds\mu}_{b_j} &=& \Biggl(\bigl(\sigma^2\tau_j^2
\bigr)^{-1}+\sum_{i=1}^n(
\xi_{ij}U_i)^T\Sigma_i^{-1}(
\xi_{ij}U_i) \Biggr)^{-1}
\\
&&{}\times  \Biggl(\sum
_{i=1}^n ({\mathbf y}_i-{\bolds
\mu}_{i(-b_j)})^T\Sigma_i^{-1}(
\xi_{ij}U_i) \Biggr)^T,
\end{eqnarray*}
and
\[
\Sigma_{b_j}= \Biggl( \bigl(\sigma^2\tau_j^2
\bigr)^{-1}+\sum_{i=1}^n (
\xi_{ij}U_i)^T\Sigma_i^{-1}(
\xi_{ij}U_i) \Biggr)^{-1},
\]
and the posterior distribution for ${\mathbf c}_j$ is $\operatorname{MVN}_v({\bolds\mu}_{c_j},\Sigma_{c_j})$, with
\begin{eqnarray*}
{\bolds\mu}_{c_j} &=& \Biggl(\bigl(\sigma^2\tau_j^{*2}
\bigr)^{-1}+\sum_{i=1}^n (
\zeta_{ij}U_i)^T\Sigma_i^{-1}(
\zeta_{ij}U_i) \Biggr)^{-1}
\\
&&{}\times  \Biggl(\sum
_{i=1}^n ({\mathbf y}_i-{\bolds
\mu}_{i(-c_j)})^T\Sigma_i^{-1}(
\zeta_{ij}U_i) \Biggr)^T,
\end{eqnarray*}
and
\[
\Sigma_{c_j}= \Biggl(\bigl(\sigma^2\tau_j^{*2}
\bigr)^{-1}+\sum_{i=1}^n(
\zeta_{ij}U_i)^T\Sigma_i^{-1}(
\zeta_{ij}U_i) \Biggr)^{-1}.
\]

Now, we derive the conditional posterior distribution for $\tau_j^2$
and $\lambda^2$ from the joint posterior distribution. Since
\begin{eqnarray*}
\nonumber
&&\pi\bigl(\tau_j^2 |{\mathbf y},{\mathbf m},{\mathbf
r}_k,{\mathbf b}_j,\lambda,{\mathbf c}_j,
\tau^{*2}_j,\lambda^*,\sigma^2,\rho\bigr)
\\
\nonumber
&&\qquad \propto\pi\bigl(\tau_j^2|\lambda\bigr)\pi
\bigl({\mathbf b}_j|\tau_j^2,
\sigma^2\bigr)
\\
\nonumber
&&\qquad \propto\bigl(\tau_j^2\bigr)^{\sklvfrac{v+1}{2}-1}
\operatorname{exp} \biggl(-\tau_j^2\frac{v \lambda^2}{2}
\biggr) \bigl(\tau_j^2\bigr)^{-\sfrac{v}{2}}
\\
&&\quad\qquad{}\times
\operatorname{exp} \biggl(-\frac{1}{2}{\mathbf b}_j^T
\bigl(\sigma^2\operatorname{diag}\bigl(\tau_j^2,
\ldots,\tau_j^2\bigr)\bigr)^{-1}{\mathbf
b}_j \biggr)
\\
&&\qquad \propto\operatorname{exp} \biggl(-\tau_j^2
\frac{v\lambda^2}{2}-\frac{1}{2\sigma^2\tau_j^2}\llVert{\mathbf
b}_j\rrVert
^2 \biggr) \bigl(\tau_j^2
\bigr)^{-1/2},
\end{eqnarray*}
and
\begin{eqnarray*}
\nonumber
&&\pi\bigl(\lambda^2|{\mathbf y},{\mathbf m},{\mathbf
r}_k,{\mathbf b}_j,\tau_j^2,{\mathbf
c}_j,\tau^{*2}_j,\lambda^*,
\sigma^2,\rho\bigr)
\\
\nonumber
&&\qquad \propto\pi\bigl(\lambda^2\bigr)\prod
_{j=1}^p\pi\bigl(\tau_j^2|
\lambda\bigr)
\\
&&\qquad \propto\bigl(\lambda^2\bigr)^{a-1}\operatorname{exp}
\bigl(-b\lambda^2 \bigr)\prod_{j=1}^p
\biggl(\frac{v\lambda^2}{2} \biggr)^{\vfrac{v+1}{2}}\operatorname{exp}
\biggl(-
\frac{v\lambda^2}{2}\tau_j^2 \biggr),
\end{eqnarray*}
the\vspace*{-2pt} posterior distribution for $\frac{1}{\tau_j^2}$ is
inverse-Gaussian$ (v\lambda^2,\sqrt{\frac{v\lambda^2\sigma^2}{\llVert
{\mathbf b}_j\rrVert ^2}} )$ and the posterior distribution for
$\lambda^2$ is Gamma$ (a+\frac{pv+p}{2},b+\frac{v\sum_{j=1}^p\tau
_j^2}{2} )$.

Similarly, the posterior distribution for $\frac{1}{\tau_j^{*2}}$ is
inverse-Gaussian$(v \lambda^{*2},\break \sqrt{\frac{v \lambda^{*2}\sigma
^2}{\llVert {\mathbf b}_j\rrVert ^2}})$, and the posterior
distribution for $\lambda^{*2}$ is Gamma$(a^*+\frac{pv+p}{2},b^*+\frac
{v\sum_{j=1}^p\tau_j^{*2}}{2})$. From these posteriors, we can see that
the hierarchical expansion of the Multivariate Laplace prior indeed
gives closed forms of posterior distributions for efficient Gibbs sampling.

Last, if we assume a stationary AR(1) covariance structure, that is,
\begin{eqnarray*}
\Sigma_i = \sigma^2 \Gamma_i =
\sigma^2 \pmatrix{ 1&\rho^{|t_{i1}-t_{i2}|}& \cdots&\rho^{|t_{iT_i}-t_{i1}|}
\vspace*{3pt}\cr
\rho^{|t_{i2}-t_{i1}|}&1& \cdots&\rho^{|t_{iT_i}-t_{i2}|}
\vspace*{3pt}\cr
\vdots&\vdots&\vdots&
\vdots
\vspace*{3pt}\cr
\rho^{|t_{i1}-t_{iT_i}|}&\rho^{|t_{i2}-t_{iT_i}|}& \cdots&1},
\end{eqnarray*}
the posterior distribution for $\sigma^2$ is an inverse chi-square
distribution, or
\[
\pi\bigl(\sigma^2|\cdot\bigr)\sim\mbox{Inv-}\chi^2
\Biggl(\sum_{i=1}^nT_i,
\frac{\sum_{i=1}^n({\mathbf y}_i-{\bolds\mu}_i)^T \Gamma
_i^{-1}({\mathbf y}_i-{\bolds\mu}_i)}{\sum_{i=1}^nT_i} \Biggr),
\]
where the first parameter is the degree of the freedom parameter and
the second one is the scale parameter, and
\begin{eqnarray*}
\nonumber
\pi(\rho|\cdot)&\propto&\pi({\mathbf y}|\cdot)\pi(\rho)
\\
&\propto&\prod_{i=1}^n\bigl(\llvert
\Gamma_i\rrvert^{-1/2}\bigr) \operatorname{exp} \Biggl(-
\frac{1}{2}\sum_{i=1}^n({\mathbf
y}_i-{\bolds\mu}_i)^T \Gamma_i^{-1}({
\mathbf y}_i-{\bolds\mu}_i) \Biggr).
\end{eqnarray*}
Based on this expression, the corresponding Metropolis--Hastings
algorithm can be developed to update $\rho$.

We use MCMC algorithms to estimate the posterior distribution of each
parameter by drawing posterior samples from the corresponding
conditional posterior distribution, given the current values of all
other parameters and the observed data. We use the potential scale
reduction factor [PSRF; \citet{GelRub92}; \citet{Geletal04}]
to access the convergence. Squared PSRF is defined as the ratio of the
marginal posterior variance to the within-chain variance, and a PSRF
less than 1.1 indicates good convergence. We run 4000 additional
iterations after all chains converge.

\section{Computer simulation}\label{sec5}
We first investigate the new Bayesian group lasso approach for
selecting important time-varying effects through simulation studies.
We generate data in the \textit{f}GWAS setting according to the model (\ref
{eqregressionGWAS}) with the number of covariates $q=1$, the number of
SNPs $p = 3000$, and the number of individuals $n=600$ or $800$.
Following the simulation techniques in the literature, genotypical data
$\xi_{ij}$ is derived from $u_{ij}$ for $i=1,\ldots,n$ and $j=1,\ldots
,p$, where each $u_{ij}$ has a standard normal distribution marginally,
and $\operatorname{cov}(u_{ij},u_{ik}) = \rho_G = 0.1$ or 0.5, representing two levels
of linkage disequilibrium. We set
\[
\xi_{ij}= \cases{ 1, &\quad$u_{ij}>c$,
\vspace*{3pt}\cr
0, &\quad$-c\le
u_{ij}\le c$,
\vspace*{3pt}\cr
-1,&\quad$u_{ij}<-c$,}
\]
where $c$ is used to determine the minor allele frequencies. Then, we
derive the indicator matrix $\zeta_{ij}$ of dominant effects from $\xi_{ij}$.

We assume that the dynamic pattern of the trait is controlled by 5 SNPs
and 1 covariate. In particular, we set ${\mathbf b}_j={\mathbf0}$ for
$j=4,\ldots,p$, and ${\mathbf c}_j={\mathbf0}$ for $j=1,2,6,\ldots,p$.
Sex is included as a covariate and is generated by randomly assigning a
sex to each subject. The time-varying effects of overall mean,
covariate and causal SNPs are generated by Legendre polynomials, with
Legendre coefficients listed in Table~\ref{tab1}. The true polynomial degrees
for these causal SNPs could be 0, 1, 2 or 3, allowing constant genetic
effects, linear genetic effects or more complicated patterns of genetic control.

\begin{table}
\tabcolsep=0pt
\caption{Parameters used in the simulated example}\label{tab1}
\begin{tabular*}{\tablewidth}{@{\extracolsep{\fill}}@{}lcd{2.2}d{2.2}d{2.2}d{2.2}@{}}
\hline
&&\multicolumn{4}{c@{}}{\textbf{Legendre coefficients}}\\[-6pt]
&&\multicolumn{4}{c@{}}{\hrulefill}\\
\textbf{Time-varying effect} & \textbf{Parameter} & \multicolumn{1}{c}{\textbf{0}} &
\multicolumn{1}{c}{\textbf{1}} & \multicolumn{1}{c}{\textbf{2}} &\multicolumn{1}{c}{\textbf{3}}\\
\hline
Mean effect&${\mathbf m}$&13.40&-3.08&1.88&-3.20\\[3pt]
Covariate effect&${\mathbf r}_1$&3.00&0.15&-2.67&3.25\\[3pt]
Additive effect&${\mathbf b}_1$&1.04&0.88&-2.05& 0.00\\
&${\mathbf b}_2$&1.17&-0.22&0.74&-4.72\\
&${\mathbf b}_3$&1.40& 0.00 & 0.00 & 0.00 \\[3pt]
Dominant effect&${\mathbf c}_3$&1.49&-2.13&4.82&1.42\\
& ${\mathbf c}_4$&1.00&1.32&1.90&1.50\\
&${\mathbf c}_5$&1.26&-1.22& 0.00 & 0.00\\
\hline
\end{tabular*}
\end{table}

To simulate irregular longitudinal phenotypical data, we assume that
the number of measurements for each subject is between 5 and 12, and
all subjects are in the age range of 30 to 80 years. For each subject
with a specific number of measurements, traits of interest are observed
at ages randomly drawn from 30 to 80.
The residual covariance matrix among different time points was assumed
to be AR(1) with $\rho=0.4$ and $\sigma^2=4, 9$ or $16$. The\vadjust{\goodbreak} phenotypes
observed at subject-specific time points and genotypes of all subjects
are collected for Bayesian analysis.

For each simulated data set, we implement MCMC algorithms as described
in Section~\ref{sec4}. In practice, the degree of Legendre polynomials should be
determined {a priori}. We recommend a procedure that analyzes all
SNPs with different polynomial degrees, where group lasso penalties are
used to regularize the estimation. When the polynomial degree is 0
(constant effect), the group lasso penalty reduces to a lasso penalty.
Then the polynomial degree $\hat{v}$ that gives the lowest Bayesian
information criterion (BIC) of the final model is chosen. In
simulations, however, this is computationally expensive. Therefore, the
polynomial degree is fixed at $\hat{v}=3$ in simulation studies.
Simulation results (see Table~\ref{tab2}) suggest
that, as long as the specified polynomial degree is greater than or
equal to the largest degree of all nonzero effects, the proposed
framework works well in selecting casual SNPs and estimating their
time-varying effects.

\begin{table}
\tabcolsep=0pt
\caption{Variable selection performance in the simulated example\vspace*{-11pt}}\label{tab2}
\begin{tabular*}{\tablewidth}{@{\extracolsep{\fill}}@{}lccccccc@{}}\\
\hline
&&\multicolumn{2}{c}{\textbf{No. of nonzeros}} &\multicolumn{3}{c}{\textbf{Proportion of}}\\[-6pt]
&&\multicolumn{2}{c}{\hrulefill} &\multicolumn{3}{c}{\hrulefill}\\
\textbf{n} & $\bolds{\sigma^2}$& \textbf{C} & \textbf{IC} &\textbf{Under-fit} &\textbf{Correct-fit}&\textbf{Over-fit} & \textbf{Time (h)}\\
\hline
\multicolumn{8}{@{}l@{}}{$\rho_G = 0.1$}\\
600 & 16 & 3.77 & 0.00 & 0.86 & 0.14 & 0.00 & 17.99 \\
600 & \phantom{0}9 & 4.93 & 0.00 & 0.07 & 0.93 & 0.00 & 17.67 \\
600 & \phantom{0}4 & 5.00 & 0.00 & 0.00 & 1.00 & 0.00 & 17.40 \\
800 & 16 & 4.99 & 0.00 & 0.01 & 0.99 & 0.00 & 23.69 \\
800 & \phantom{0}9 & 5.00 & 0.00 & 0.00 & 1.00 & 0.00 & 24.78 \\
800 & \phantom{0}4 & 5.00 & 0.00 & 0.00 & 1.00 & 0.00 & 24.35
\\[6pt]
\multicolumn{8}{@{}l@{}}{$\rho_G = 0.5$}\\
600 & 16 & 4.61 & 0.00 & 0.35 & 0.65 & 0.00 & 17.90 \\
600 & \phantom{0}9 & 5.00 & 0.00 & 0.00 & 1.00 & 0.00 & 17.29 \\
600 & \phantom{0}4 & 5.00 & 0.00 & 0.00 & 1.00 & 0.00 & 17.63 \\
800 & 16 & 5.00 & 0.00 & 0.00 & 1.00 & 0.00 & 23.97 \\
800 & \phantom{0}9 & 5.00 & 0.00 & 0.00 & 1.00 & 0.00 & 23.49 \\
800 & \phantom{0}4 & 5.00 & 0.00 & 0.00 & 1.00 & 0.00 & 24.38 \\
\hline
\end{tabular*}
\end{table}

Once all posterior samples are collected from MCMC algorithms, SNPs are
selected in the following way: a time-varying additive effect $a_j(t)$
or dominant effect $d_j(t)$ is included in the final model if at least
one of its four Legendre coefficients has a two-sided 95\% interval
estimate that does not cover zero. In the supplemental article [\citet{Lietal15}], we plot the potential scale reduction factor against
iterations for each parameter in ${\mathbf b}_1$, ${\mathbf b}_2$,
${\mathbf b}_3$, ${\mathbf c}_3$, ${\mathbf c}_4$ and ${\mathbf c}_5$.
This is a simulation randomly drawn from the specification $n = 600$
and $\sigma^2=16$. All chains converge very quickly and stay below the
threshold of 1.05 (the red line).

To evaluate the variable selection performance of the proposed
procedure, we calculate several measures of model sparsity for the
final model, which are summarized in Table~\ref
{tab2}. Column ``C'' shows the average number of
SNPs with nonzero varying-coefficients correctly included in the final
model, and column ``IC'' is the average number of SNPs with no genetic
effect incorrectly included in the final model. Column ``Under-fit''
represents the proportion of excluding any relevant SNP in the final
model. Similarly, column ``Correct-fit'' represents the proportion that
the extract true model was selected and column ``Over-fit'' gives the
proportion of including all relevant SNPs as well as one or more
irrelevant SNPs. Clearly, both sample size and the noise level play
important roles in how well the Bayesian group lasso could select the
exactly correct model. However, as sample size decreases and noise
increases, our procedure tends to select fewer important SNPs rather
than produce more false positives. Moreover, the impact of linkage
disequilibrium is limited, and our method works slightly better in the
presence of high linkage disequilibrium.

Other than the performance of selecting truly important SNPs, we
further investigate how well the procedure estimates the time-varying
effects of selected SNPs. To ameliorate the bias of the parameter
estimates introduced by group lasso penalties, we always refit the \textit{f}GWAS model after variable selections, where only selected SNPs are
included in the final model and all regularization parameters are set
to zero. For each time-varying genetic effect of important SNPs,
Tables~1~and~2 in the supplemental article [\citet{Lietal15}]
summarize the average estimates, standard errors and the mean squared
errors (MSEs) of Legendre coefficients over replications where the
effect is selected for $\rho_G = 0.1$. As can be seen from these
tables, both bias and standard error decrease as noise level decreases.
MSEs are slightly lower for additive effects and lower order Legendre
coefficients.

To compare the parameter estimates with those produced by another
strategy aimed at the same genetic model, we implement the univariate
\textit{f}GWAS approach by \citet{Dasetal11} using the same data set.
Specifically, this single-SNP analysis extends the traditional GWAS
analysis framework by allowing the phenotype to be collected repeatedly
over time and approximating the time-varying genetic effects by
Legendre polynomials. A Benjamini-Hochberg false discovery rate (FDR)
controlling procedure is used to adjust for multiple comparisons in
selecting significant SNPs. Table~3 in the supplemental article [\citet{Lietal15}] shows that this single-SNP analysis produces biased
estimates for all parameters.\footnote{Since this approach cannot
identify if the significance is due to the additive effect or the
dominant effect, both effects are reported for five important SNPs.}

Finally, we compare the variable selection performance of four
approaches: (1)~a~Bayesian group lasso; (2) a univariate \textit{f}GWAS
approach by \citet{Dasetal11}; (3) a functional principal component
analysis (fPCA) approach [\citet{RamSil05}] that analyzes
the fPCA of the\vspace*{1pt} longitudinal phenotype; and (4)~a~slope model that
simplifies the longitudinal phenotype to its slope.\footnote{We thank
the Associate Editor and an anonymous referee for pointing out the fPCA
method and the slope method, respectively.} In the third and the fourth
model, the leading three fPCA scores and the slope calculated from each
growth curve are tested against genetic predictors, respectively, where
group lasso or lasso regressions with 5-fold cross-validation are used
to select relevant SNPs.

\begin{table}
\tabcolsep=0pt
\caption{Variable selection performance of alternative methods in the simulated example}\label{tablevariableselectionDas&Li}\label{tab3}
\begin{tabular*}{\tablewidth}{@{\extracolsep{\fill}}@{}lccccccccccc@{}}
\hline
&&\multicolumn{2}{c}{\textbf{Nonzeros}}&\multicolumn{3}{c}{\textbf{Proportion of}}
&\multicolumn{2}{c}{\textbf{Nonzeros}}&\multicolumn{3}{c@{}}{\textbf{Proportion of}}
\\[-6pt]
&&\multicolumn{2}{c}{\hrulefill}&\multicolumn{3}{c}{\hrulefill}
&\multicolumn{2}{c}{\hrulefill}&\multicolumn{3}{c@{}}{\hrulefill}
\\
\textbf{n}& $ \bolds{\sigma^2}$ &\textbf{C}& \textbf{IC} &\textbf{U.-fit}&\textbf{C.-fit}&\textbf{O.-fit}& \textbf{C}&\textbf{IC}&\textbf{U.-fit}&\textbf{C.-fit}&\textbf{O.-fit}\\
\hline
\multicolumn{7}{@{}l}{\textit{Bayesian group lasso}} & \multicolumn{5}{l@{}}{\textit{\citeauthor{Dasetal11}} (\citeyear{Dasetal11})}\\
600 & 16 & 3.93 & \phantom{0}0.00 & 0.83 & 0.17 & 0.00 & 4.94 & \phantom{0}0.51 & 0.06 & 0.55 & 0.39 \\
600 & \phantom{0}9 & 4.80 & \phantom{0}0.00 & 0.20 & 0.80 & 0.00 & 5.00 & \phantom{0}0.20 & 0.00 & 0.83 & 0.17\\
600 & \phantom{0}4 & 5.00 & \phantom{0}0.00 & 0.00 & 1.00 & 0.00 & 4.98 & \phantom{0}0.02 & 0.02 & 0.96 & 0.02\\
800 & 16 & 4.86 & \phantom{0}0.00 & 0.14 & 0.86 & 0.00 & 4.99 & \phantom{0}0.40 & 0.01 & 0.70 & 0.29 \\
800 & \phantom{0}9 & 5.00 & \phantom{0}0.00 & 0.00 & 1.00 & 0.00 & 5.00 & \phantom{0}0.23 & 0.00 & 0.81 & 0.19\\
800 & \phantom{0}4 & 5.00 & \phantom{0}0.00 & 0.00 & 1.00 & 0.00 & 5.00 & \phantom{0}0.01 & 0.00 & 0.99 & 0.01
\\[3pt]
\multicolumn{7}{@{}l}{\textit{Functional PCA}} & \multicolumn{5}{l@{}}{\textit{Slope model}}\\
600 & 16 & 0.51 & \phantom{0}4.46 & 1.00 & 0.00 & 0.00 & 1.06 & 10.28 & 1.00 & 0.00 & 0.00 \\
600 & \phantom{0}9 & 1.19 & \phantom{0}7.01 & 0.98 & 0.00 & 0.02 & 2.53 & 16.42 & 0.99 & 0.00 & 0.01 \\
600 & \phantom{0}4 & 2.75 & 13.53 & 0.78 & 0.00 & 0.22 & 3.65 & 21.22 & 0.97 & 0.00 & 0.03 \\
800 & 16 & 0.77 & \phantom{0}5.02 & 1.00 & 0.00 & 0.00 & 1.82 & 10.91 & 1.00 & 0.00 & 0.00 \\
800 & \phantom{0}9 & 1.79 & 11.14 & 0.88 & 0.00 & 0.12 & 3.06 & 14.90 & 0.98 & 0.00 & 0.02 \\
800 & \phantom{0}4 & 2.90 & 17.16 & 0.61 & 0.00 & 0.39 & 3.82 & 19.00 & 0.99 & 0.00 & 0.01 \\
\hline
\end{tabular*}
\end{table}

For fairness of comparison, longitudinal phenotype data are not
generated from our nonparametric genetic model (\ref
{eqregressionGWAS}). Instead, we use the same genotype data with $\rho
_G = 0.1$ but assume the following time-varying genetic effects:
$a_1(t) = 0.5 + \sin(0.2t)$, $a_2(t) = 1/(0.5 + \exp(-0.06t)) - 0.5$,
$a_3(t) = \log(0.05t)$, $d_3(t) = -1.5$, $d_4(t) = 60/t$, and $d_5(t) =
0.2 - 0.035t$ for the first five SNPs. These functional forms are
unknown to researchers. Table~\ref{tablevariableselectionDas&Li}
presents variable selection results, where all measures strongly prefer
the Bayesian group lasso. Among the alternative approaches, the
univariate \textit{f}GWAS approach has the best variable selection
performance. For the fPCA approach and the slope approach, the
probability of selecting casual SNPs increases with the signal-to-noise
ratio (column ``C''), but the proportion of under-fit is always
substantial. Interestingly, as signal-to-noise ratio increases, the
probability of identifying false positives also increases
(columns ``IC'' and ``Over-fit''), especially when $\sigma^2$ decreases from 16
to 9. The inconsistency of these procedures suggests the risk of
inflated false positive rates when only the major movements of growth
curves are captured and tested in association studies.

In the above simulation studies, the minor allele frequency is set to
0.3. Unreported simulations also demonstrate that as the minor allele
frequency decreases, both statistical powers and false positive rates
decrease. But our method is still much better than the alternative
approaches. Despite the Bayesian framework's theoretical advantages in
handling parameter uncertainty, practically it could be slower than
frequentist methods. When $n = 600$, $\sigma^2 = 9$, $\rho_G = 0.1$ and
the number of SNPs $p=1000$, the Gibbs sampler's computational time is
about 5.70 hours. Experiments show that a linear regression
line\footnote{We thank the Editor for sharing the idea of using this
regression.} can describe almost perfectly the relationship between the
computational time in hours and $p$: $\log_{10}(\mathit{time}) = 0.754 +\log
_{10}(p/1000)$.



\section{Worked example}\label{sec6}
We use the newly developed model to analyze a real GWAS data set from
the Framingham Heart Study (FHS), a cardiovascular study based in
Framingham, Massachusetts, supported by the National Heart, Lung, and
Blood Institute, in collaboration with Boston University [\citet{DawMeaMoo51}].
Recently, 550,000 SNPs have been genotyped for the entire
Framingham cohort [\citet{Jaq07}], from which 493 males and 372
females were randomly chosen for our data analysis. These subjects were
measured for body mass index (BMI) at multiple time points from age 29
to age 61. The number of measurements for a subject ranges from 2 to
18, and the intervals of measurement are also highly variable among
subjects. As is standard practice, SNPs with rare allele frequency $<$10\% were excluded from data analysis. The numbers and percentages of
nonrare allele SNPs vary among different chromosomes and range from
4417 to 28,771 and from 0.64 to 0.72, respectively.

\begin{figure}[b]

\includegraphics{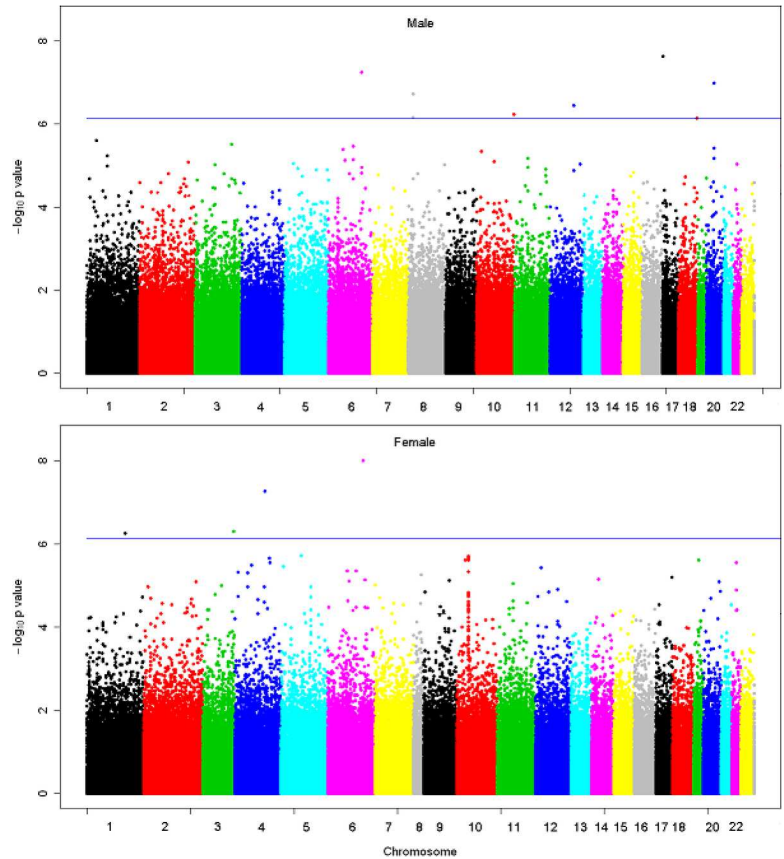}

\caption{Manhattan plot of p-values for association by genomic position
for male and female, where different colors across the $x$-axis represent
different chromosomes, and the horizontal line indicates the
significance level obtained by the Benjamini-Hochberg FDR adjustment at
$\alpha= 5\%$.}\label{fig1}
\end{figure}

A single-SNP analysis was used to analyze the phenotypic data of BMI
for males and females separately. Figure~\ref{fig1} gives $-\log_{10}$ $p$-values for each SNP in the two sexes, from which 33,239 SNPs with
$-\log_{10}$ $p$-values greater than 2.0 in at least one sex were
selected. Before applying Bayesian group lasso analysis to this
irregular longitudinal data set, we imputed missing genotypes for a
small proportion of SNPs according to the distribution of genotypes in
the population. Then, by treating the sex as a covariate, we imposed
group lasso penalties on both additive effects and dominant effects in
hopes of identifying SNPs with notable effects on BMI, where all
effects are possibly functions of time. According to our discussions in
Section~\ref{sec5}, the whole procedure was repeated with polynomial degrees: 0,
1, 2, 3 and 4, and the corresponding BICs of the final model are as
follows: 27,470, 27,444, 27,416, 27,408 and 27,426. Therefore, a polynomial
degree of 3 is appropriate in this real data example.

\begin{figure}

\includegraphics{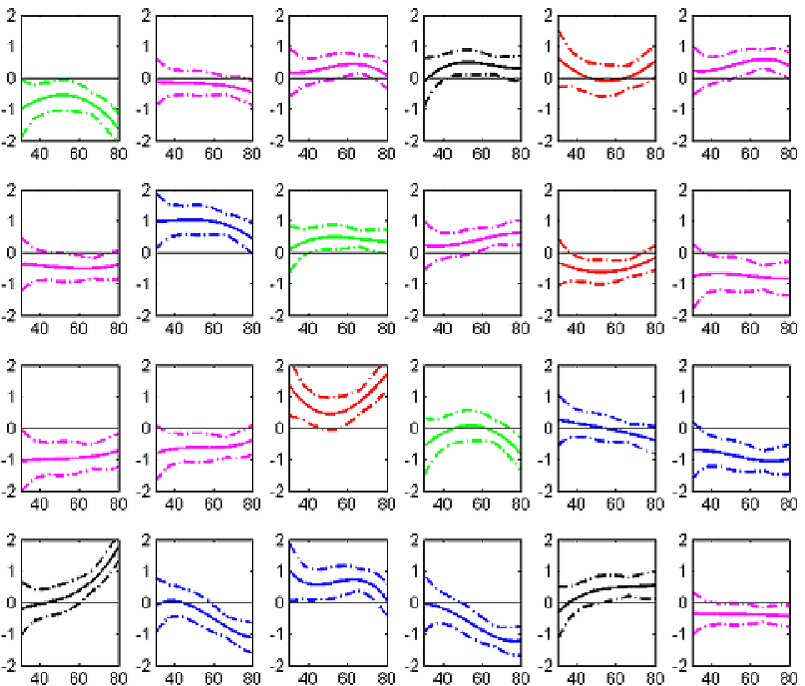}

\caption{Additive effects of selected SNPs in the real data example.}\label{fig2}
\end{figure}

\begin{figure}[b]

\includegraphics{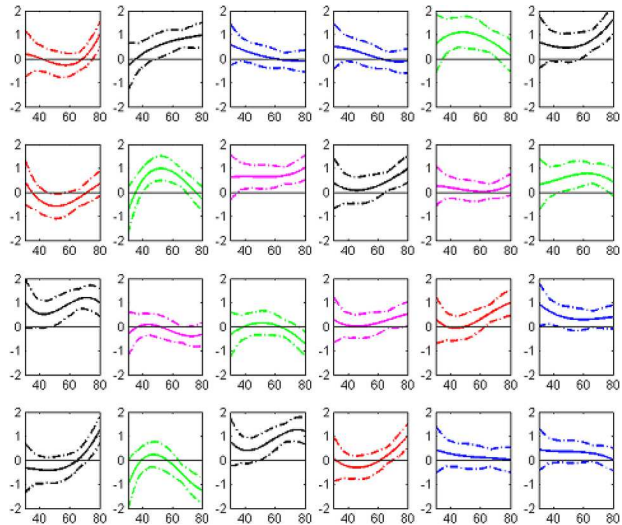}

\caption{Dominant effects of selected SNPs in the real data example.}\label{fig3}
\label{figrealeffects}
\end{figure}

The Bayesian group lasso selected 24 significant SNPs, located on
chromosomes 1, 2, 3, 4, 6, 7, 12, 14, 16 and 23. Table~\ref{tab4} tabulates the
names, positions, alleles and estimated Legendre coefficients of these
SNPs. The first allele in the column ``Alleles'' represents the minor
allele. Using the Legendre coefficient estimates, we plot their
time-varying additive effects and dominant effects in Figures~\ref{fig2}~and~\ref{fig3},
respectively, where the associated interval estimates\footnote{Suppose
for one varying-coefficient, the interval estimate of the $q$th
Legendre coefficient is $(b_{q,U}, b_{q,L})^T, q = 1, \ldots, 4$, and
the Legendre polynomials are $(u_0, u_1, u_2, u_3)^T = (1, t, \frac
{1}{2}(3t^2-1), \frac{1}{2}(5t^3-3t) )^T$ for each standardized time
point $t \in[-1,1]$. Then the interval estimate of the
varying-coefficient at time $t$ is $(\sum_{q=1}^4 \tilde{b}_{q,U} u_q,
\sum_{q=1}^4 \tilde{b}_{q,L} u_q)^T$, where $\tilde{b}_{q,U} = b_{q,U}$
if $u_q$ is positive and $b_{q,L}$ otherwise, and $\tilde{b}_{q,L} =
b_{q,L}$ if $u_q$ is positive and $b_{q,U}$ otherwise.} are also
provided. Some of these detected SNPs are located in a similar region
of candidate genes for obesity. For example, the detected SNPs on
chromosomes 4, 6 and 12 are close to candidate genes for BMI-related
type 2 diabetes [\citet{Fra07}].

%
%

Figures~\ref{fig2}~and~\ref{fig3} show that the time courses of the genetic effects of
some SNPs are relatively constant (magenta), monotonically increasing
(black) or decreasing (blue). That is, given a population carrying one
of these SNPs in the same environment, the expected BMI is different at
different ages. Individuals carrying certain SNPs may have lower BMI in
mid-life but tend to have higher BMI when they are younger or older
(red). Conversely, individuals carrying certain SNPs tend to have
higher BMI in mid-life (green), which may increase the risk for stroke
later in life, according to a prospective study [\citet{Jooetal04}].


\begin{sidewaystable}
\tabcolsep=0pt
\tablewidth=\textwidth
\caption{Information about selected SNPs in the real data example}\label{tab4}
\begin{tabular*}{\tablewidth}{@{\extracolsep{\fill}}@{}lcd{10.0}cd{2.3}d{2.3}d{2.3}d{2.3}d{2.3}d{2.3}d{2.3}d{2.3}@{}}
\hline
&&&&\multicolumn{8}{c}{\textbf{Estimated legendre coefficients}}\\[-6pt]
&&&&\multicolumn{8}{c}{\hrulefill}\\
\textbf{Chr}&\textbf{Name}& \multicolumn{1}{c}{\textbf{Position}} &
\multicolumn{1}{c}{\textbf{Alleles}} & \multicolumn{4}{c}{\textbf{Additive effect}} & \multicolumn{4}{c@{}}{\textbf{Dominant effect}}\\
\hline
\phantom{0}1 & ss66334458 & 79{,}393{,}823 & C/T & -1.504 & -1.656 & -1.334 & -0.143 & 1.136 & 2.645 & 2.480 & 0.780 \\
\phantom{0}1 & ss66050888 & 93{,}240{,}623 & A/G & -0.431 & -0.532 & -0.344 & -0.111 & 0.737 & 0.745 & -0.185 & 0.190 \\
\phantom{0}1 & ss66275851 & 93{,}245{,}738 & C/T & -0.128 & -0.949 & -1.096 & -0.448 & 0.079 & -0.254 & 0.278 & -0.004 \\
\phantom{0}1 & ss66048018 & 115{,}427{,}398 & A/G & 0.396 & 0.217 & 0.028 & 0.418 & 0.213 & -0.007 & 0.571 & 0.294 \\
\phantom{0}1 & ss66287256 & 221{,}051{,}934 & G/A & 0.497 & 0.788 & 0.934 & -0.047 & 0.386 & -1.065 & -0.672 & 0.221 \\
\phantom{0}1 & ss66104828 & 234{,}701{,}498 & A/C & 0.111 & -0.620 & -0.951 & -0.461 & 1.445 & 1.833 & 1.307 & 0.205 \\
\phantom{0}2 & ss66484730 & 103{,}489{,}666 & G/A & -0.341 & 0.254 & 0.335 & 0.098 & 0.057 & 0.612 & 0.552 & -0.565 \\
\phantom{0}2 & ss66232775 & 103{,}493{,}541 & T/C & 0.476 & -1.098 & -0.806 & -0.220 & 0.043 & -0.816 & -1.011 & 0.810 \\
\phantom{0}2 & ss66185516 & 239{,}065{,}169 & G/T & 0.397 & 0.074 & -0.053 & 0.228 & 1.039 & 0.852 & 0.687 & 0.269 \\
\phantom{0}3 & ss66397464 & 73{,}251{,}862 & C/T & 0.415 & 0.183 & -0.198 & -0.192 & 0.677 & 0.895 & 0.437 & -0.212 \\
\phantom{0}4 & ss66402098 & 186{,}281{,}132 & T/C & -0.225 & 0.630 & 0.565 & -0.009 & 0.418 & 0.651 & 0.744 & 0.244 \\
\phantom{0}6 & ss66218814 & 3{,}311{,}818 & C/T & -0.724 & 0.043 & 0.159 & 0.182 & 0.237 & -0.795 & -1.056 & -0.336 \\
\phantom{0}7 & ss66083459 & 89{,}430{,}534 & T/G & -0.744 & 0.518 & 0.336 & 0.141 & 0.377 & -1.070 & -1.555 & -1.214 \\
12 & ss66288005 & 29{,}860{,}263 & A/G & -0.342 & 0.724 & 0.541 & 0.322 & 0.096 & 0.563 & 0.875 & 0.806 \\
14 & ss66282595 & 24{,}339{,}998 & G/A & 1.461 & 1.471 & 1.217 & -0.246 & -0.588 & -1.194 & -0.973 & 0.022 \\
14 & ss66411959 & 24{,}340{,}175 & G/A & -0.782 & -1.357 & -1.311 & -0.080 & 0.307 & 0.323 & 0.068 & -0.276 \\
14 & ss66416767 & 24{,}348{,}496 & G/T & -0.232 & -0.589 & -0.117 & -0.033 & 0.507 & 0.610 & -0.098 & -0.559 \\
14 & ss66281419 & 77{,}702{,}561 & G/A & -0.802 & 0.151 & 0.488 & 0.252 & 0.438 & -0.109 & 0.254 & -0.189 \\
16 & ss66091573 & 57{,}829{,}089 & C/T & 1.402 & 2.674 & 1.450 & 0.400 & 0.999 & 2.747 & 1.859 & 0.426 \\
16 & ss66242525 & 57{,}935{,}351 & C/T & -0.548 & -0.516 & 0.465 & 0.579 & -0.537 & -0.479 & 0.093 & 1.234 \\
16 & ss66489647 & 57{,}938{,}934 & A/G & -0.217 & -2.058 & -1.834 & -1.059 & 0.595 & -0.350 & -1.070 & -0.967 \\
16 & ss66444701 & 82{,}976{,}515 & C/G & -0.672 & -0.259 & 0.831 & 0.481 & 0.639 & 1.552 & 0.876 & -0.143 \\
16 & ss66529263 & 84{,}383{,}030 & G/T & 0.478 & 0.539 & -0.033 & 0.301 & 0.066 & -0.313 & -0.037 & -0.111 \\
23 & ss66369851 & 121{,}966{,}143 & G/T & -0.419 & -0.108 & -0.052 & -0.025 & 0.050 & -0.663 & -0.454 & -0.185 \\
\hline
\end{tabular*}
\end{sidewaystable}


\section{Discussion}\label{sec7}
When the number of predictors $p$ is much larger than the number of
observations $n$, highly regularized approaches, such as penalized
regression models, are favorable to identify nonzero coefficients, to
enhance model \mbox{predictability} and to avoid overfitting [\citet{HasTibFri09}]. In this article, we proposed a Bayesian regularized estimation
procedure for nonparametric varying-coefficient models that could
simultaneously estimate time-varying effects and implement variable
selection. The procedure extends the standard Bayesian lasso [\citet{ParCas08}] and standard group lasso [\citet{YuaLin06}] to a
nonparametric setting, and is applicable to irregular longitudinal data.

We approximated time-varying effects by Legendre polynomials and
presented a Bayesian hierarchical model with group lasso penalties that
encourages sparse solutions at the group level. The group lasso
penalties are introduced by assigning multivariate Laplace priors to
regression coefficients, and are implemented on the basis of its
hierarchical expansion which yields an efficient Gibbs sampler in the
MCMC estimation. Although computationally intensive, it outperforms the
standard group lasso in the sense that it provides not only point
estimates but also interval estimates of all parameters. In addition,
the Bayesian group lasso treats the regularization parameters as
unknown hyperparameters and estimates them along with other parameters.
This technique avoids choosing the tuning parameters by
cross-validation and automatically accounts for the uncertainty in its
selection that affects the estimates of regression coefficients.

In one of the most powerful but challenging areas in genetics, we
incorporated our new procedure to genome-wide association studies
(GWAS) by testing a large number of SNPs simultaneously, particularly
with $p\gg n$, based on the dynamic pattern of genetic effects on
complex phenotypes or diseases. We first applied the new approach to
\textit{f}GWAS for age-specific changes of BMI and successfully identified
several significant SNPs, some of which are confirmed by empirical
genetic studies [\citet{Fra07}]. For example, previous molecular
studies have observed a candidate gene (\textit{FTO}) coding
alpha-ketoglutarate-dependent dioxygenase, a fat mass and
obesity-associated protein. Our model detected SNPs ss66091573,
ss66242525 and ss66489647 on chromosome 16 in a region of the FTO gene,
suggesting the biological relevance of these SNPs in fat-related trait
control. Our model also detected other SNPs in close proximity of
different candidate genes; that is, SNP ss66397464 in peroxisome
proliferator-activated receptor-$\gamma$ gene (PPARG) on chromosome 3,
SNP ss66402098 in the Wolfram syndrome 1 gene (WFSI) on chromosome 4,
SNP ss66218814 in CDK5 regulatory-subunit-associated protein 1-like 1
gene (CDKAL1) on chromosome 6, and SNP ss66288005 in potassium
inwardly-rectifying channel, subfamily J, member 11 gene (KCNJ11) on
chromosome 12 [\citet{Fra07}]. Among these four genes, PPARG and KCNJ
were found to be associated with obesity [\citet{Videtal97};
\citet{Moretal10}], while WFSI and CDKAL1 are believed to be
associated with diabetes [\citet{Sanetal07}; \citet{Scoetal07};
\citet{Steetal07}]. Therefore, all these discoveries have
well validated the biological relevance of the new model.

To address challenges for the post-GWAS era, genetic association
studies began to focus on SNPs within a set of functional candidate
genes. For instance, \citet{Micetal10} analyzed 566 SNPs from 14
candidate genes that are believed to be associated with asthma. \citet{XuTay09} developed tools to recommend SNPs based on information on
gene expression studies, regulatory pathways and functional regions
that appear to be linked to the disease. In their example, 1361 SNPs
were recommended for a genetic association study on prostate cancer.
These tools could be used as a preprocessing step for the proposed
procedure in this article. Statistically, on the other hand, variable
screening approaches [\citet{FanLv08}] for longitudinal data can be
developed to recommend a subset of SNPs.

From a theoretical point of view, the proposed method can also
approximate varying-coefficients by nonparametric techniques other than
Legendre\vadjust{\goodbreak} polynomials, and model the within-subject correlation by other
parametric or nonparametric covariance structures. Given its potential
influence, an optimal model for longitudinal covariance structure
should be chosen based on the nature of practical data [\citet{Zhaetal05}; \citet{YapFanWu09}]. More generally, it can be easily extended
to the problem where the number of variables in each group varies, such
as the multi-factor ANOVA with each factor having several levels. Also,
gene-gene interactions and gene-environment interactions can be
incorporated to better decipher a detailed picture of the genetic
architecture of a complex trait.


\section*{Acknowledgments}
Jiahan Li and Zhong Wang
contributed equally to this work.
The authors are grateful to the Editor, the Associate Editor and three
anonymous referees for providing valuable comments that significantly
improved the paper.

\begin{supplement}[id=suppA]
\stitle{Convergence diagnostics and summary of parameter estimates}
\slink[doi]{10.1214/15-AOAS808SUPP} 
\sdatatype{.pdf}
\sfilename{aoas808\_supp.pdf}
\sdescription{We plot the potential scale reduction factor (PSRF)
against iterations and summarize the average estimates, standard errors
and mean squared errors (MSEs) of corresponding Legendre coefficients
for the first five genetic predictors.}
\end{supplement}

%

\printaddresses
\end{document}